\documentclass[prl,aps,twocolumn,showpacs]{revtex4}
 \newcommand{\mytitle}[1]{
 \twocolumn[\hsize\textwidth\columnwidth\hsize
 \csname@twocolumnfalse\endcsname #1 \vspace{1mm}]}
 \newcommand{\beq}{\begin{equation}}
 \newcommand{\eeq}{\end{equation}}
 \newcommand{\bea}{\begin{eqnarray}}
 \newcommand{\eea}{\end{eqnarray}}
 \newcommand{\pdag}{{\phantom{\dagger}}}
 
\usepackage{graphics}
\usepackage{graphicx}
\usepackage{amssymb}
\usepackage{bm}
\begin{document}

\title{Tunable Fano-Kondo resonance in side-coupled double quantum dot system}
\author{Chung-Hou Chung$^{1,2}$, Tsung-Han Lee$^{1}$}
\affiliation{
$^{1}$Electrophysics Department, National Chiao-Tung University,
HsinChu, Taiwan, 300, R.O.C. \\
$^{2}$Department of Physics and Applied Physics, Yale University, New Haven, CT 06520, USA \\
}

\date{\today}

\begin{abstract} 
We study the interference between the Fano and 
Kondo effects in a side-coupled double-quantum-dot system where
one of the quantum dots couples to conduction electron bath while
the other dot only side-couples to the first dot via antiferromagnetic
(AF) spin exchange coupling. We apply both the 
perturbative renormalization group 
(RG) and numerical renormalization group (NRG) approaches to 
study the effect of AF coupling 
on the Fano lineshape in the conduction leads. With particle-hole symmetry, 
the  AF exchange coupling competes with the Kondo effect 
and leads to a local spin-singlet ground state for arbitrary small coupling, 
so called ``two-stage Kondo effect''. 
As a result, via NRG we find the spectral properties 
of the Fano lineshape in the  
tunneling density of states (TDOS) $\rho_c(\omega)$ 
of conduction electron leads shows double dip-peak features  
at the energy scale around the Kondo temperature and the one much below it,  
corresponding to the two-stage Kondo effect; it also shows an 
universal scaling behavior at very low energies. 
We find the qualitative agreement between 
the NRG and  the perturbative RG approach. 
Relevance of our work to the experiments is discussed.
 
\end{abstract}

\maketitle

\begin{center}
{\bf I. Introduction} 
\end{center}
 Fano resonance is the quantum interference effect between a localized
state with finite-width and a conduction band\cite{fano}. 
The hallmark of the Fano 
resonance is the asymmetric lineshape in tunnelling density of states (TDOS) 
of the conduction band.  One example of Fano resonance
is the transport through low dimensional electronic (Fermi) system with 
local impurities. The Kondo effect\cite{kondo} plays an important role if
these impurities carry unpaired spins.
 Recently, there has been growing interest both 
theoretically and experimentally  
in the Fano resonance associated with the Kondo effect via the 
STM measurements of noble metal surfaces\cite{xiang, kroha, schiller, 
gadzuk, wingreen, schneider, eigler} 
as well as in quantum dot devices\cite{hofstetter, sato}. 
The Fano resonance in these systems in general arises from two quantum 
interference effects: 1. between the broadened local level and the 
continuum conduction band and 2. between the Kondo resonance in 
the local level and the conduction band. The combined two effects 
give rise to rather complicated lineshpe in STM measurement of the TDOS. 
The Fano resonance in TDOS of conduction electrons in such systems 
can be served as an alternative approach to study the Kondo effect 
in addition to the local density of states of the quantum dot.
The Fano lineshape in TDOS of conduction electrons in the leads 
of a single Kondo dot system 
has been extensively studied, and it is sensitive to both 
the spatial phase of the free conduction electrons and the scattering phase 
shift associated with the Kondo effect.

Very recently, the Fano resonance has been extended 
experimentally\cite{sasaki} and theoretically\cite{lee, nishi} to 
the side-coupled 
double quantum dot system where the competition between Kondo and Fano 
effects gives rise to change in conductance profile. In this paper, we 
investigate the Fano-Kondo interference in the side-coupled 
double-quantum-dot systems where only one of the two dots (dot $1$) connects 
to the leads while the other isolated dot 
(dot $2$) is side-coupled to dot $1$\cite{side, grempel}. 
In the Kondo limit where charging 
energy on each dot is large, an antiferromagnetic (AF) spin exchange (RKKY) 
coupling is generated via the second-order hoping between two dots 
competes with the Kondo effect, leading to local spin-singlet ground 
state for arbitrary finite values of $J$, so-called "two-stage Kondo 
effect"\cite{side,grempel}. 
Previous studies on the side-coupled double-dot systems have been 
mostly focused on the dip of LDOS on dot $1$ upon applying the AF 
RKKY coupling. However, little is known about the feedback effect of 
the two-stage Kondo effect mentioned above on the TDOS of 
conduction electrons in the leads. In this paper, we generalize the 
Fano lineshape in TDOS of electrons in the leads as a result of the 
two-stage Kondo effect in side-coupled double-quantum-dot system. 
The systematic perturbative and numerical renormalization group 
approaches are applied here in the cases both with and without 
particle-hole symmetry. We find as a consequence of the two-stage Kondo 
effect, the spectral property of the 
Fano lineshape in TDOS of the leads develops an asymmetrical double 
dip/peak structure; it also shows an universal scaling behavior at 
very low energies. 
We compare our NRG results with the perturbative RG analysis.

\begin{center}
{\bf II. The Model Hamiltonian.}
\end{center}
 Our starting  Hamiltonian for the side-coupled 
double-dot system is the single-impurity Anderson model for dot $1$ 
with additional antiferromagnetic spin-exchange coupling between dot $1$ 
and the isolated dot $2$ which side-coupled to it\cite{side}. 
\begin{eqnarray}
H &=& \sum_{k,\alpha=L,R} \epsilon_k c^{\dagger}_{k\alpha\sigma} c_{k\alpha\sigma} + 
\sum_{\alpha=L,R}\sum_{k,\sigma } ( t_{\alpha} c^{\dagger}_{k \alpha \sigma} d_{1,\sigma} + h.c.)\nonumber \\
   &+& \sum_{i\sigma} \epsilon_{d i} d_{i\sigma}^{\dagger} d_{i\sigma} + \sum_{i=1,2} U_i n_i^{\uparrow} n_i^{\downarrow}\nonumber \\
 &+& J \, {\bf S}_1\, {\bf S}_2,
\end{eqnarray}
where $t_L$ and $t_R$   denote the tunneling amplitudes to  the left and right
leads, respectively, and  $c^{\dagger}_{\alpha \epsilon \sigma}$ creates an electron 
in lead $\alpha=L,R$ with spin $\sigma$. 
This tunnel coupling leads to a broadening of the level on dot 1, the width of
which is given  by  $\Gamma =\Gamma_{L}+\Gamma _{R}= 2\pi (t_L^{2}\varrho_{L} + t_R^{2}\varrho_{R})$, with  
$\varrho_{L/R}$ the density  of states in the leads. 
Here, $i=1,2$ labels the two dots, and ${\bf S}_{i}=(1/2)\sum_{\sigma \sigma ^{\prime}}d_{i\sigma }^{\dagger }{\bf \sigma}_{\sigma \sigma
^{\prime }}d_{i\sigma ^{\prime }}^{\pdag}$ is their spin. 
Each dot is subject to a charging energy, $U_1\approx U_2 = U = E_C$. 
In the presence of particle-hole symmetry, we have $\epsilon_{d i} = -\frac{U_i}{2}$.
Note that in the Kondo limit where the charging energy $E_C$ is large, the direct 
hoping between the two dots are strongly suppressed and an antiferromagnetic spin exchange coupling $J>0$ is generated 
via the second-order hoping processes. 

The physical observables of our interest are: (i). 
LDOS of impurity on dot $1$: 
$\rho_{d1}(\omega) = \frac{-1}{\pi} Im G_{d1}(\omega)$ 
and (ii) the TDOS of the conduction electron $\rho_c(\omega)$: 
$\rho_c(\omega) = \rho_0 + \delta \rho_c(\omega)$ 
where $\rho_0$ is the density of states of the bare conduction electron leads 
: $\rho_0=\frac{-1}{\pi}ImG_c^0(\omega=0)$ with 
$G_{c}^0(\omega -{\it i}\eta)$ being the bare conduction electron 
Green's function, and $\delta \rho_c(\omega)$ is 
the correction to the LDOS of the conduction electron due to the coupling 
between leads and the quantum dot system:  
$\delta \rho_c(\omega) = \frac{-1}{\pi} 
Im \delta G_c(\omega - {{\it i} \eta})$. Here, the correction 
to the  
conduction electron Green's function 
$\delta G_c(\omega - {\it i}\eta)$ is given by:
\begin{equation}
\delta G_{c}(\omega  - {\it i}\eta) 
= \frac{\Gamma}{\pi\rho_0} G_{c}^0(\omega -{\it i}\eta) G_{d1}(\omega -{\it i} \eta)  
G_{c}^0(\omega -{\it i}\eta)
\label{deltaGc}
\end{equation}
Using Eq. \ref{deltaGc}, we have\cite{kroha}
\begin{eqnarray}
& & \delta \rho_c(\omega) = -\Gamma \rho_0 \times \nonumber \\
& & [(q_c^2 -1) Im G_{d1}(\omega - {\it i} \eta) 
- 2 q_c Re G_{d1}(\omega -{\it i}\eta)], 
\label{deltarhoc}
\end{eqnarray}
with $q_c$ being defined as
\begin{equation}
q_c = -\frac{Re G_{c}^0(\omega - {\it i} \eta)}
{Im G_{c}^0(\omega - {\it i} \eta)}, 
\end{equation}
and it can be treated approximately as an frequency-independent constant
\cite{kroha,xiang}. Following Ref. \cite{side}, below we apply both 
perturbative renormalization group (RG) 
and numerical renormalization group (NRG) approaches to 
calculate these quantities in the presence of particle-hole 
symmetry. Though the LDOS on dot $1$ 
(or equivalently the imaginary part of the Green's function 
on dot $1$, $ImG_{d1}(\omega)$) at finite RKKY coupling $J$ 
via both RG and NRG has been computed in Ref. \cite{side}, the real 
part of $G_{d1}(\omega)$, $ReG_{d1}(\omega)$, which is also needed 
to analyze the spectral 
property of the Fano lineshape in the TDOS of the conduction electron leads 
($\rho_c(\omega)$), 
has not yet been calculated by either perturbative RG or NRG approach. 
In the following, we provide a numerical and analytical 
analysis on the Fano lineshape for $\rho_c(\omega)$  
 by analyzing both the real and the imaginary parts of 
$G_{d1}(\omega)$ at finite $J$ via NRG and compare them with those 
via perturbative RG approach.

First, we discuss the case for $J=0$. 
For $J=0$ and in the presence of particle-hole symmetry 
($\epsilon_{di}=-U_i/2$), it has been known that in the Kondo regime  
$\omega \ll T_k$ 
with $T_k\approx D_0 e^{-\pi \Gamma /U_1}$ being the Kondo temperature 
for dot $1$, 
$G_{d1}(\omega)$ is well approximated by the single 
Lorentzian\cite{side}: 
\begin{equation}
G_{d1}(\omega) \approx T_{d1}^{0}(\omega - {\it i} \eta)=
\frac{z}{\omega   +{\it i}\;\tilde T_K + {\it i} \eta }\;,
\label{Gd10}
\end{equation}
with $z= c\; \frac{T_{K}}{\Gamma }$ being the quasi-particle weight at the
Fermi energy, and $\tilde T_K = z\Gamma = c\; T_K$ being an energy of the 
order of the Kondo temperature, $T_K$. 
The precise value of the universal constant $c$ relating $T_K$ and $\tilde
T_K$  depends on the definition of $T_K$. Here, we define $T_K$ as the
half-width of the transmission $T(\omega)\equiv -\Gamma ImG_{d1}(\omega)$. 
From fitting $G_{d1}(\omega)$ with the NRG data, we get $c\approx 0.5$. 
Note that by Fermi-liquid theory and 
principles of 
renormalization group, the RKKY interaction also 
gets renormalized by the same $z-$ factor: 
$J\rightarrow \frac{\tilde{J}} z$\cite{side}. 
Here, $\tilde J$ is slightly different from $J$ due to the large logarithmic 
tail in $ImG_{d1}(\omega)$. The value of $\tilde{J}$ is 
obtained from the fit of $ImG_{d1}(\omega)$ to NRG data: 
$\tilde{J}\approx 1.1 J$\cite{side}. 
However, for $\omega\ge T_k$, 
the above simple Lorentzian approximation fails to account 
for the large logarithmic tail in $ImG_{d1}(\omega)$. 
Therefore, corrections to the single Lorentzian 
approximation are needed in this case 
to more accurately describe 
$G_{d1}(\omega)$. 
Via the Dyson equation approach, taking into account the 
interference between the Kondo 
resonance and the broadened impurity level, 
we obtain a more accurate description for 
the Green's function of the dot $1$\cite{xiang}:
\begin{equation}
G_{d1}(\omega) = G_{d1}^0(\omega) + G_{d1}^0(\omega) 
\tilde{T}_{d1}(\omega) G_{d1}^0(\omega)
\label{Gd1J0}
\end{equation}
where the bare Green's function on dot $1$, $G_{d1}^0(\omega)$, 
describing a local impurity level with a level broadening $\Gamma$ and 
LDOS $\rho_{d0}\equiv \frac{-1}{\pi} ImG_{d1}^0(\omega=0)$, is given by:
\begin{equation}
G_{d1}^0 = \frac{1-n/2}{\omega - \epsilon_{d1} + {\it i} \Gamma} + 
 \frac{n/2}{\omega - \epsilon_{d1} - U_1 + {\it i} \Gamma} 
\end{equation}
with $n=<n_{d1}^{\uparrow} + n_{d1}^{\downarrow}>$ being the average occupation number on dot $1$; and $\tilde{T}_{d1}(\omega)$ is the scattering $T-$matrix corresponding to the Kondo resonance, given approximately by\cite{xiang}: 
\begin{equation}
\tilde{T}_{d1}(\omega - {\it i} \eta)\approx 
\frac{b e^{{\it i} 2\delta}}{\omega -\epsilon_K  +{\it i}\;\tilde T_K + {\it i} \eta }\; 
\label{Td10}
\end{equation}
with $b$ being a fitting parameter to be fitted with the NRG data for $ImG_{d1}(\omega)$. In the presence of particle-hole symmetry, we have  
$n=1$, $\epsilon_K=0$. Here, $\delta$ in Eq. \ref{Td10} corresponds to 
the phase shift associated with the Kondo resonance scattering, 
and it gives $\delta=\pi/2$ in the case of particle-hole symmetry. 
By fitting Eq. \ref{Gd1J0} with the NRG data, we 
find $b\approx z/ (\pi \rho_{d0})^2$, which is in good agreement 
with the known result: $-ImG_{d1}(\omega=0)= 1/\Gamma$ for a single impurity 
Anderson model\cite{kondo}. 
In the Kondo regime ($\omega \ll T_k$ and $E_c\gg \Gamma$) 
of our system and for $J=0$, the bare Green's function 
on dot $1$, $G_{d1}^0(\omega)$, are approximately given by: 
$ReG_{d1}^0(\omega)\approx 0$, 
$\frac{-ImG_{d1}^0(\omega)}{\pi}\approx \rho_{d0}$. 
The above approximations 
lead to the following approximated expressions for 
$G_{d1}(\omega)$ after including the interference between the 
Kondo resonance and the broadened impurity (dot $1$) level via 
Eq. \ref{Gd1J0}:
\begin{eqnarray}
ReG_{d1}(\omega)&\approx& (\pi \rho_{d0})^2\frac{b\omega}{\omega^2+\tilde{T}_K^2}\nonumber \\
ImG_{d1}(\omega)&\approx& -\pi \rho_{d0} -(\pi\rho_{d0})^2
\frac{b\tilde{T}_K}
{\omega^2+\tilde{T}_K^2}
\label{ReImGdapprox}
\end{eqnarray}
with $\rho_{d0} = -\frac{1}{\pi} ImG_{d1}^0(\omega=0)$ being 
the LDOS of dot $1$ at $\omega=0$. 
From Eq. \ref{deltarhoc} and Eq. \ref{ReImGdapprox}, 
in the Kondo limit 
the correction to conduction electron density of states can therefore 
be expressed in terms of the well-known Fano lineshape\cite{xiang,kroha}:
\begin{equation}
\delta\rho_c(\omega, J=0)\approx \rho_0 (\frac{q_c^2+2q_c\epsilon-1}{\epsilon^2+1}+\beta),
\label{Fano_form}
\end{equation}
where $\epsilon=\frac{\omega -\epsilon_K}{\tilde{T}_K},\beta=\pi \rho_{d0}\Gamma(q_c^2-1)$.
Note that in general the Dyson equation approach in 
Eq. \ref{Gd1J0} is also valid for both 
$\omega\ll T_k$ and $\omega\approx T_k$  
in the presence of large particle-hole asymmetry: 
$|\epsilon_{d1}-\epsilon_F|\le \Gamma$ (with $\epsilon_F$ being the 
Fermi energy of the leads) where the the interference between the Kondo and 
broadened impurity level plays an important role in 
$G_{d1}(\omega)$\cite{xiang}.\\
\begin{figure}[t]
\begin{center}
\includegraphics[width=8.0cm]{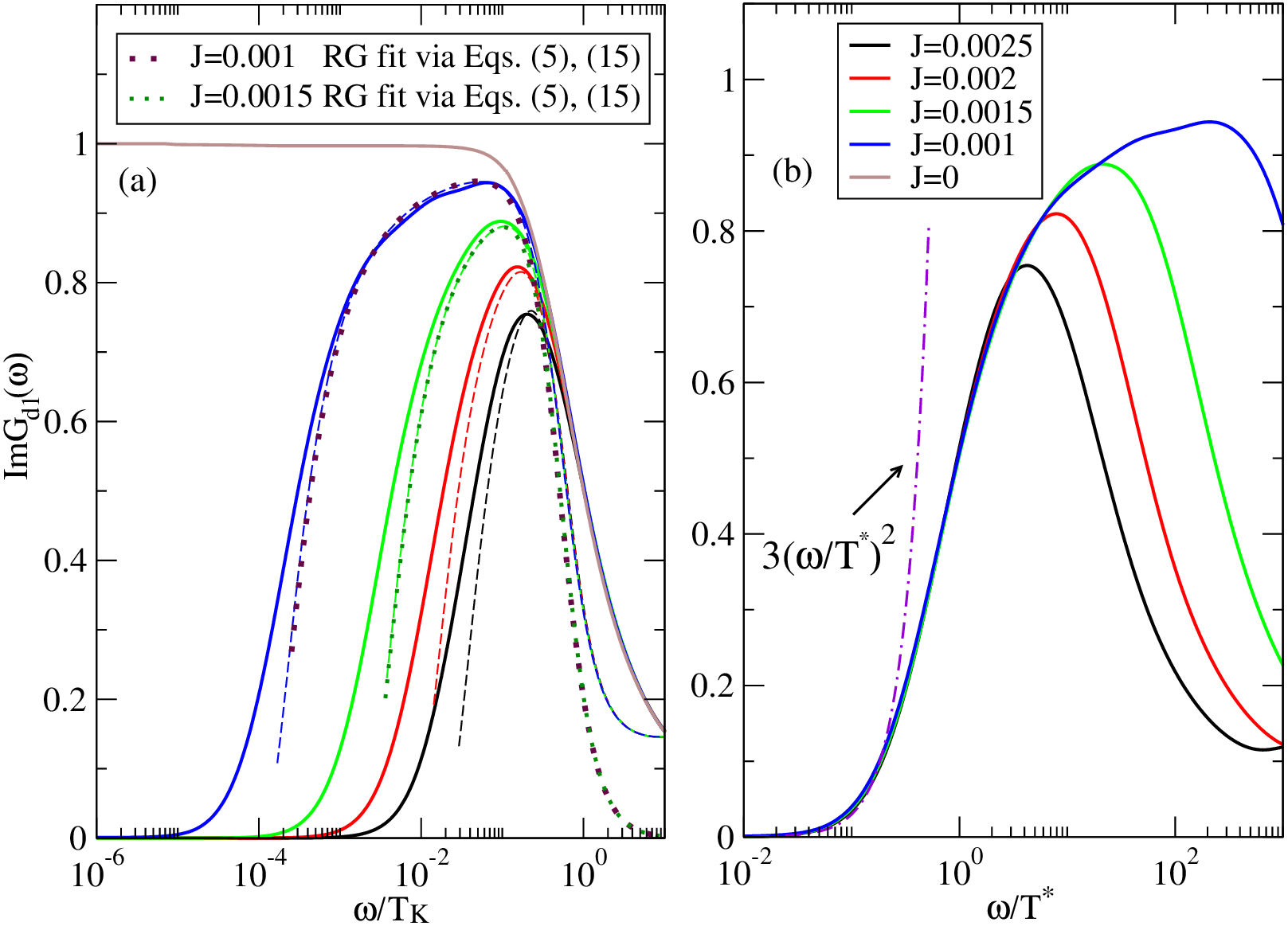}
\end{center}
\par
\vspace{-0.0cm}
\caption{(a). $ImG_{d1}(\omega)$ (normalized to 
$ImG_{d1}(\omega=0, J=0)\approx -1/\Gamma$) 
of dot $1$ with particle-hole symmetry 
for different antiferromagnetic RKKY couplings $J$ 
calculated by NRG (solid lines) and perturbative RG (dashed lines) 
via Eqs. \ref{Gd1J0} and \ref{G_d1J} (dashed lines). 
Dotted lines are RG fits via Eqs. \ref{Gd10} and \ref{G_d1J}. 
The NRG parameters are $U_1=U_2=D_0$, $\epsilon_{d1}=\epsilon_{d2}=-0.5 D_0$, 
$\Gamma=0.2 D_0$ with $D_0=1$. For $J=0$, we find $T_k\approx 0.005 D_0$. 
The fitting parameters $c\approx 0.5$, and $\tilde{J}\approx 1.1$. (b). $ImG_{d1}(\omega/T^*)$ 
(normalized to $ImG_{d1}(\omega=0, J=0)\approx -1/\Gamma$) shows an universal 
scaling behavior for $\omega < T^*$.  
The dot-dash line is the power-law $(\omega/T^*)^2$ fit to 
the crossover function of $ImG_{d1}(\omega/T^*)$ for $\omega\ll T^*$, see 
Eq. \ref{ImGd_scaling}.}
\label{LDOS}
\end{figure}

\begin{center}
{\bf III. Perturbative Renormalization Group analysis} 
\end{center}
Now, we turn on a finite RKKY coupling $J$.  
Following Ref. \cite{side}, to gain an analytical understanding we employing 
the perturbative renormalization group analysis in the limit of 
$J\rightarrow 0$. We restrict ourselves the case with particle-hole symmetry.  
Though some of the aspects in this case 
has been studied in Ref. \cite{side}, 
it proves to be useful to summarize 
its key results for further calculations on the Fano lineshape for 
$\rho_c(\omega)$ in the presence of RKKY coupling $J$. 
In the limit  $J\to 0$, ``two-stage Kondo
screening'' takes place\cite{side,grempel}: The spin of dot $1$ first 
gets Kondo screened 
below  Kondo temperature $T_{K}\approx
D\;e^{-\pi U/\Gamma }$ with $D$ being the bandwidth cutoff, the first stage 
Kondo effect. Then for energy scale 
much below $T_k$, the second stage Kondo effect occurs at
 $\omega <T^*\ll T_k$ between dot $1$ and $2$ via 
the antiferromagnetic RKKY coupling $J$ where the spin on the dot $2$ gets 
Kondo screened. Here, the Kondo resonance peak in electron 
density of states on dot $1$ plays the effective fermionic bath for the 
second stage Kondo effect. We will discuss how the Fano lineshape for 
$\rho_c(\omega)$ is affected 
in the presence of the antiferromagnetic RKKY coupling. 
Summing up all leading logarithmic vertex diagrams leads to the 
following scaling equation for the dimensionless vertex function\cite{side}
\begin{equation} 
\frac{d (\gamma (\omega, \tilde{T_k}))}{dl}\equiv \frac{d (\hat  \varrho(\omega) \tilde J)}{dl} = ({\hat  \varrho}(\omega) \tilde J)^2\;,
\label{scaling}
\end{equation}
with the scaling variable defined as $l\equiv \log (\tilde T_K / \tilde
T_K')$. Here, 
$\hat  \varrho(\omega)\equiv \varrho(\omega)/z = \frac{-1}{\pi z}ImG_{d1}(\omega)$ is the rescaled effective density of states of dot $1$.  
Integrating this differential equation up to 
$l\equiv \log (\tilde T_K / \omega)$,  one obtains the dimensionless vertex 
function in the leading logarithmic approximation\cite{side}:  
\begin{equation} 
\gamma(\omega,\tilde T_K) = 
\frac1 {\frac {\omega^2}{\tilde T_K^2} \log\frac {\tilde T_K}{T^*}
+ \log\frac {|\omega|}{T^*}}\;,
\label{gamma}
\end{equation}
with the second scale $T^*$ defined as 
\begin{equation}
T^* = \tilde T_K \; \exp(-\pi\;\tilde T_K/\tilde J)\;.
\label{eq:Tstar}
\end{equation}
The second order self-energy correction to the retarded Green's 
function $G^0_{d1}$ simply gives the expression\cite{side} 
\begin{equation}
\Sigma (\omega )= S(S+1)\frac {\tilde J ^{2}}{4 z^2} G_{d1}(\omega)
\end{equation}
where S=1/2. 
The Green's function of dot $1$ after including self-energy and vertex 
correction is given by\cite{side}:
\begin{equation}
G^{J}_{d1}(\omega) = \frac{z}
{z G_{d1}^{-1}(\omega) - 
 \frac{\tilde{J}^2(\omega) S(S+1)}{4z} G_{d1}(\omega)}. 
\label{G_d1J}
\end{equation}
where $\tilde{J}(\omega)$ is replaced by 
$\gamma(\omega)/\hat \varrho(\omega)$, 
and $G_{d1}(\omega)$ is given by either Eq. \ref{Gd10} (the Dyson equation 
approach) or 
Eq. \ref{Gd1J0} (the single Lorentzian approximation). 
Note that due to the logarithmic corrections in $\gamma$, the 
spectral density of dot $1$ develops a dip at energies 
$\omega\sim T^\ast \ll T_K$ for any infinitesmall $J$, which 
suppresses the low-energy transmission coefficient through dot $1$. 
Physically, this comes from as a consequence of the 
fact that electrons of energy 
$\omega< T^\ast$ are not energetic enough to break up the local 
spin singlet and therefore their transport is suppressed. 
For a finite RKKY coupling $J>0$, 
the real and imaginary parts of $G_{d1}^J(\omega)$ obtained in Eq. 
\ref{G_d1J} via perturbative RG approach 
lead to an analytical expression for the 
correction to the LDOS on dot $1$, $\delta \rho_c^J(\omega)$:
\begin{equation}
\delta\rho_c^J(\omega)= 
-\Gamma\rho_0[(q_c^2-1)ImG_{d1}^J(\omega)-2q_cReG_{d1}^J(\omega)].
\label{Fano_form_J}
\end{equation}
 Below we present the results via NRG with fits by the perturbative 
RG calculations.\\
 \begin{figure}[t]
\begin{center}
\includegraphics[width=8.0cm]{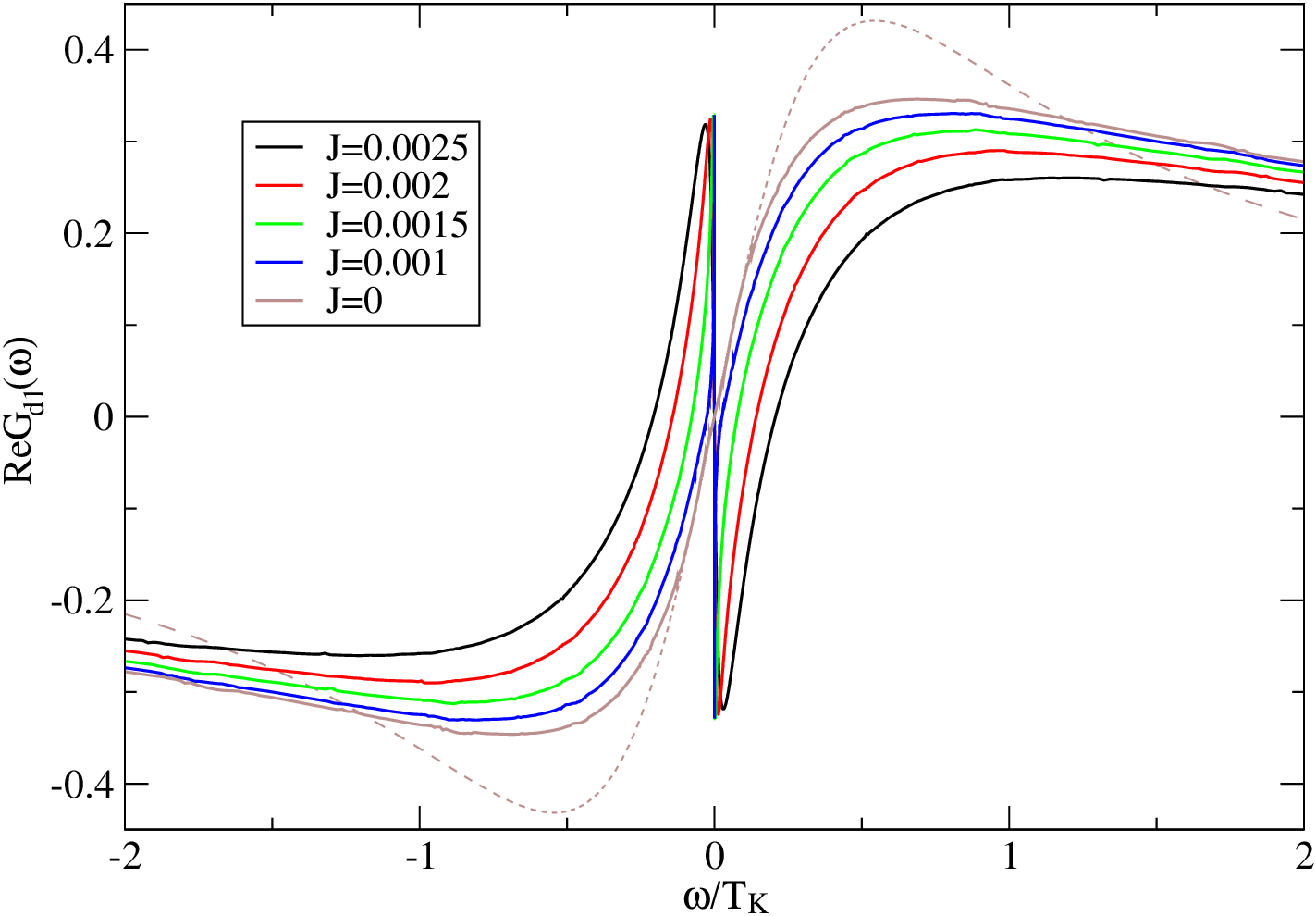}
\end{center}
\par
\vspace{-0.5cm}
\caption{$ReG_{d1}(\omega)$ (normalized to 
$-ImG_{d1}(\omega=0, J=0)$) 
of dot $1$  
for different antiferromagnetic RKKY couplings $J$ 
by NRG (solid lines). The dashed line is a fit to the NRG 
data for $J=0$ via Eq. \ref{Gd1J0}. 
The other parameters are the same as in Fig. \ref{LDOS}.}
\vspace{0.7cm}
\label{ReGd}
\end{figure}

\begin{figure}[t]
\begin{center}
\includegraphics[width=8.5cm]{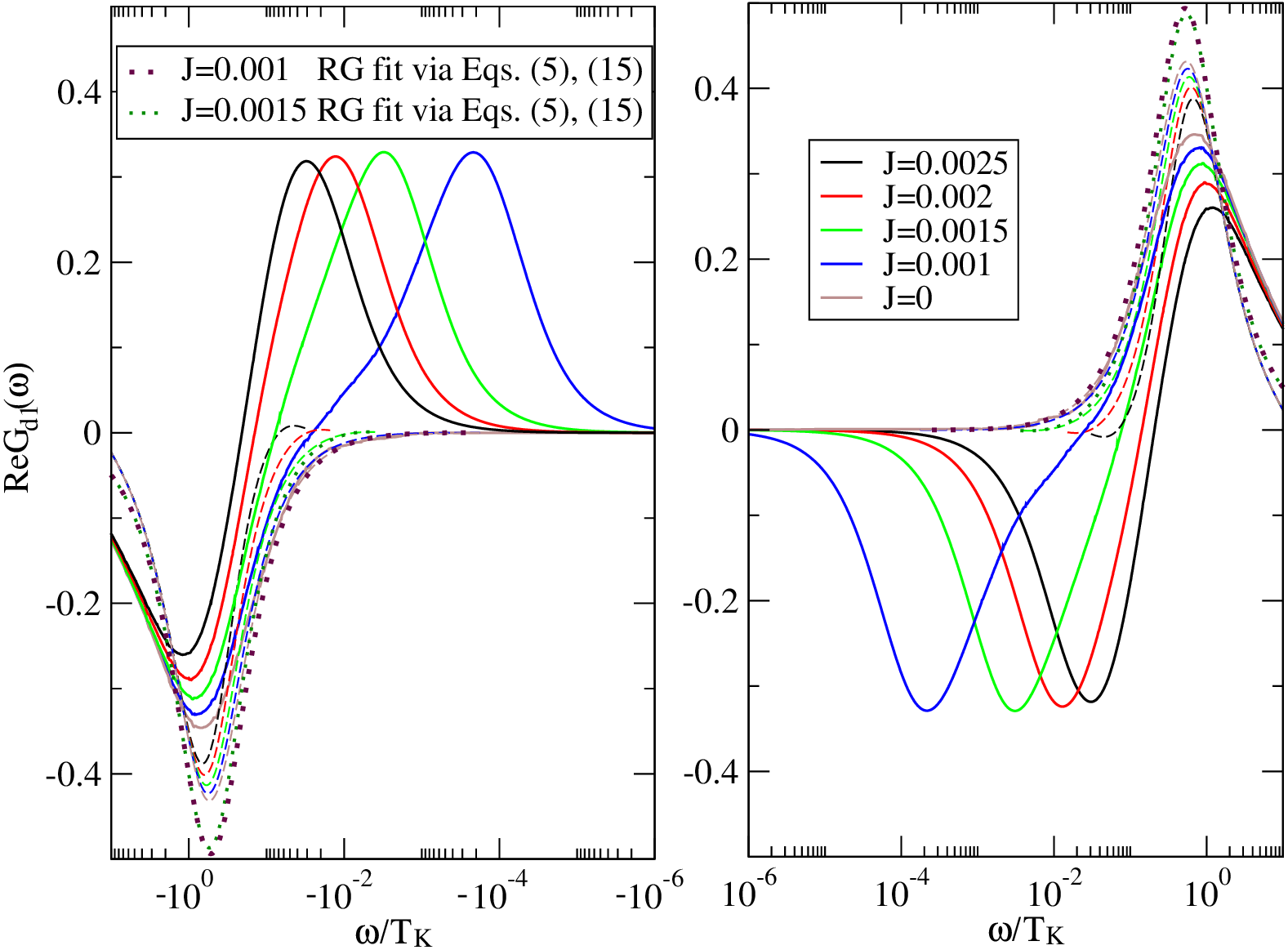}
\end{center}
\par
\vspace{-0.5cm}
\caption{$ReG_{d1}(\omega)$ (normalized to $-ImG_{d1}(\omega=0, J=0)$) on a logarithmic scale of $\omega/T_k$  
for different antiferromagnetic RKKY couplings $J$ 
by NRG (solid lines) and perturbative RG via Eqs. \ref{Gd1J0} and \ref{G_d1J} 
(dashed lines). Dotted lines are RG fits via Eqs. \ref{Gd10} and \ref{G_d1J}. The other 
parameters are the same as in Fig\ref{LDOS}.}
\vspace{0.5cm}
\label{ReGd_log}
\end{figure}

\begin{figure}[t]
\begin{center}
\includegraphics[width=8.5cm]{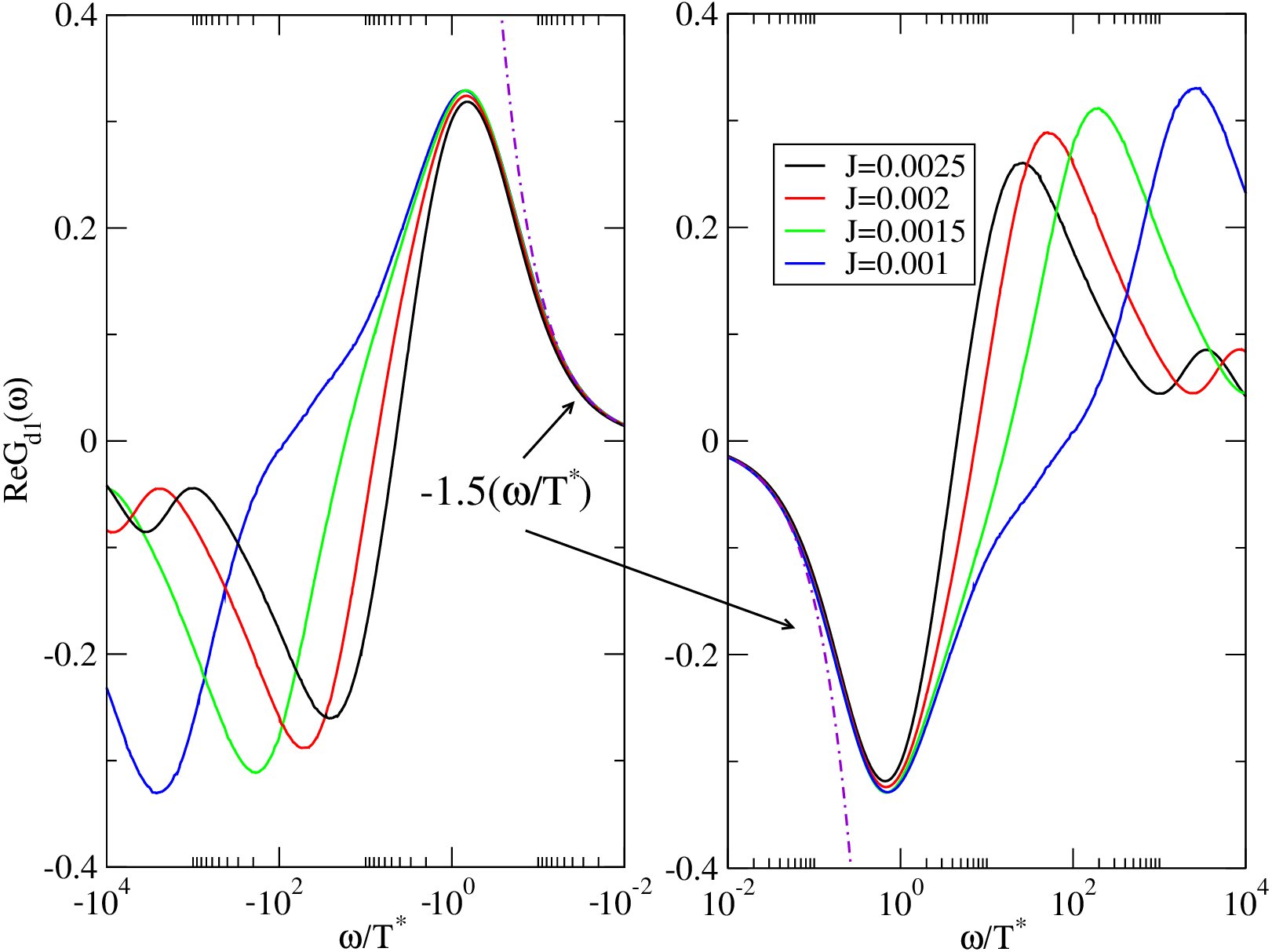}
\end{center}
\par
\vspace{-0.5cm}
\caption{$ReG_{d1}(\omega)$ (normalized to $-ImG_{d1}(\omega=0, J=0)$)  on a logarithmic scale of $\omega/T^*$  
for different antiferromagnetic RKKY couplings $J$ 
by NRG. The other parameters are the same as in Fig. \ref{LDOS}. 
 The dot-dash lines are power-law $(\omega/T^*)$ fits to the 
universal crossover function of $ReG_{d1}(\omega)$ 
for $\omega \ll T^*$, see Eq. \ref{regd_scale}.}
\vspace{0.5cm}
\label{ReGd_scaling}
\end{figure}
\begin{center}
{\bf IV. Comparison to the Numerical Renormalization Group (NRG) analysis.} 
\end{center}
We have performed the NRG calculations on the system in the 
presence of particle-hole symmetry. The NRG parameters we used 
are: $U_1=U_2=D_0=1$, $\epsilon_{d1} = \epsilon_{d2}=-0.5$, 
$\Lambda=2$, $\Gamma_L=\Gamma_R=0.1$ with $D_0$ being the 
bandwidth of the conduction electron baths. (Here, we set $D_0=1$ 
as the unit of all parameters.) Within each NRG iteration, we keep 
the lowest $1000$ states. For $J=0$, we find $T_k\approx 0.005 D_0$. 
As RKKY coupling is increased, 
both real and imaginary parts of 
$G_{d1}(\omega)$ get splited at $\omega=0$. 
First, as shown in Ref. \cite{side}, 
the imaginary part of $G_{d1}(\omega)$ (proportional to 
DOS of dot $1$) at finite $J$ 
shows a dip below the characteristic energy scale $T^*$ 
for any arbitrary $J>0$ (see Fig.\ref{LDOS}(a)). 
For small  
RKKY coupling $J$, 
the NRG results for $ImG_{d1}(\omega)$ can be fitted 
reasonably well by the perturbative RG approach over an intermediate 
energy range $T^* \ll \omega \ll T_k$. Furthermore, a clear universal 
scaling behavior of the KT type is observed from the NRG results of 
$ImG_{d1}(\omega)$ for $\omega\le T^*$: 
$ImG_{d1}(\omega)\approx g_0 g(\omega/T^*)$ (see Fig. \ref{LDOS}(b))
\cite{side}. With particle-hole symmetry, 
the scaling function $g(\omega/T^*)$ is completely universal. 
As pointed out in Ref. \cite{side}, 
the ground state of the system at any finite $J$ is 
a local spin-singlet (a Fermi liquid), the very low energy 
crossover of $ImG_{d1}(\omega)$ for $\omega\ll T_k$ 
vanishes as $(\omega/T^*)^2$, following the Fermi liquid behavior:
\begin{equation}
-\Gamma ImG_{d1}(\omega) \approx a_1 (\frac{\omega}{T^*})^{2}.
\label{ImGd_scaling}
\end{equation} 
where $a_1\approx 3.0$ from the fit to the NRG data (see Fig. \ref{LDOS} (b))
\cite{side}. Note that we find the perturbative RG approach  via 
Eq. \ref{Gd1J0} leads to a better 
fit to the NRG results 
for $ImG_{d1}(\omega)$ than that via Eq. \ref{Gd10}, as expected.\\

We now discuss the real part of $G_{d1}(\omega)$. 
For $J=0$, $ReG_{d1}(\omega)$ is antisymmetric with respect to 
$\omega=0$ and it shows a peak/dip at 
$\omega\approx \pm T_k$, signature of the first Kondo effect.  
As the RKKY coupling $J$ is increased, the magnitude of 
the peak/dip in $ReG_{d1}(\omega)$ for $\omega\approx T_k$ 
decreases, indicating the Kondo effect is suppressed. 
At a much lower energy scale, $T^*\approx\omega\ll T_k$, 
the Kondo dip-peak structure in $ReG_{d1}(\omega)$ gets a further 
 split with a width $D \approx 2 T^*$: 
it develops a negative-valued dip for $\omega \approx T^*$; while it 
shows a positive-valued peak for $\omega \approx -T^*$. 
In the 
$\omega\rightarrow 0$ limit, both positive and negative branches of 
$ReG_{d1}(\omega)$ vanish (see Fig. \ref{ReGd} and Fig. \ref{ReGd_log}). 
We can get an analytical understanding of this behavior as follows:  
In the Kondo regime $\omega \ll T_k$, the real part of $G_{d1}^{J}(\omega)$ 
is approximately given by (see Eq. \ref{G_d1J}) 
\begin{eqnarray}
ReG_{d1}^J(\omega) &\approx& \frac{z \omega 
(1-  \frac{3 \tilde{J}^2(\omega) }{16 \tilde{T}_k^2})}
{\omega^2 (1-  \frac{3 \tilde{J}^2(\omega) }{16 \tilde{T}_k^2})^2  + 
 \tilde{T}_k^2(1-  \frac{3 \tilde{J}^2(\omega) }{16 \tilde{T}_k^2})^2}
\nonumber \\ 
\label{analyticReGd}
\end{eqnarray}
\begin{figure}[t]
\begin{center}
\includegraphics[width=8.0cm]{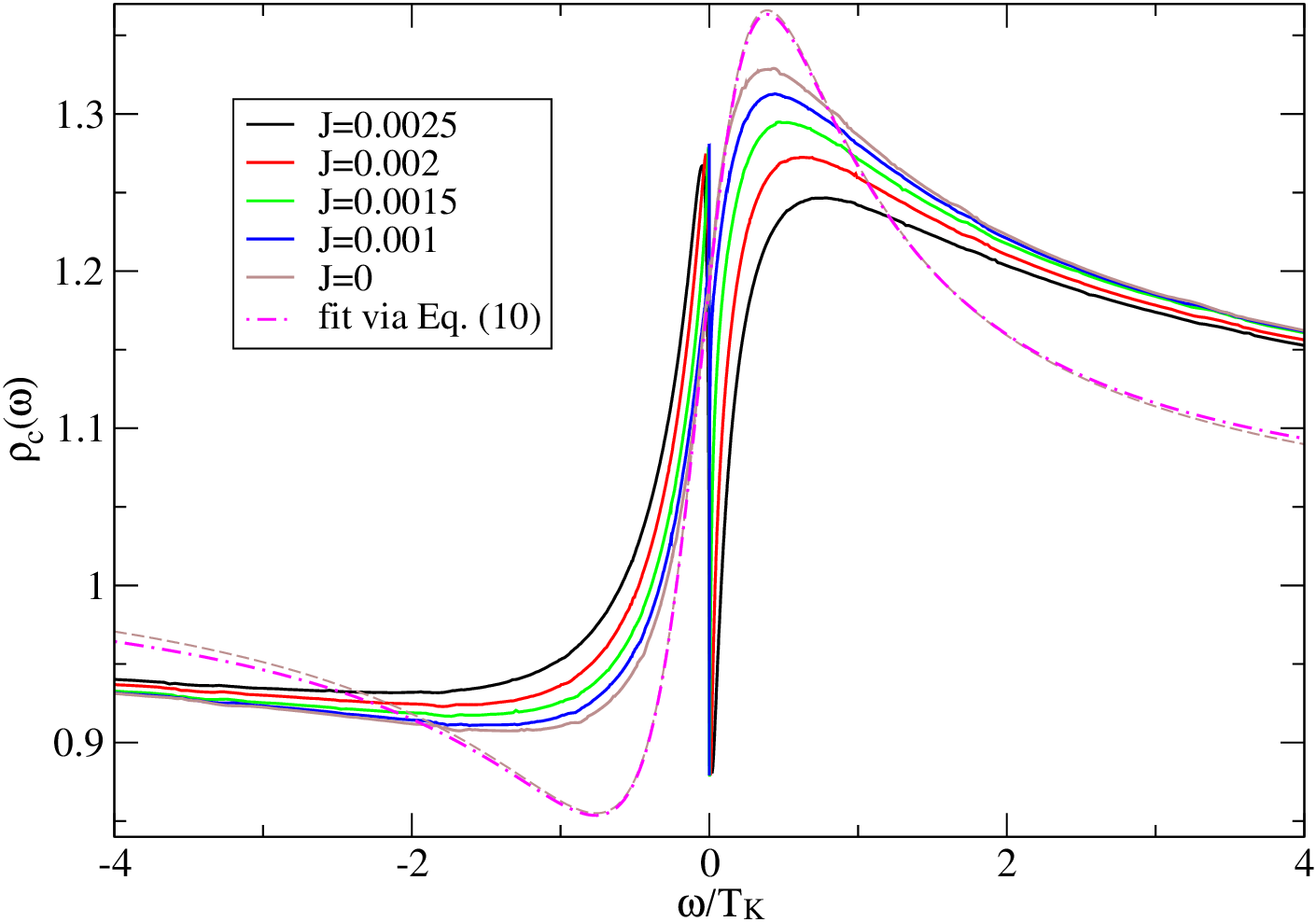}
\end{center}
\par
\vspace{-0.5cm}
\caption{The Fano lineshape for $\rho_c(\omega)$ 
(in unit of $\rho_0$)
for different antiferromagnetic RKKY couplings $J$ 
by NRG. The dashed line is a fit to the Fano lineshape 
form Eq. \ref{Fano_form} for J=0.  
The other parameters are the same as in Fig. 1.}
\vspace{0.7cm}
\label{Fano}
\end{figure}
\begin{figure}[t]
\begin{center}
\includegraphics[width=8.5cm]{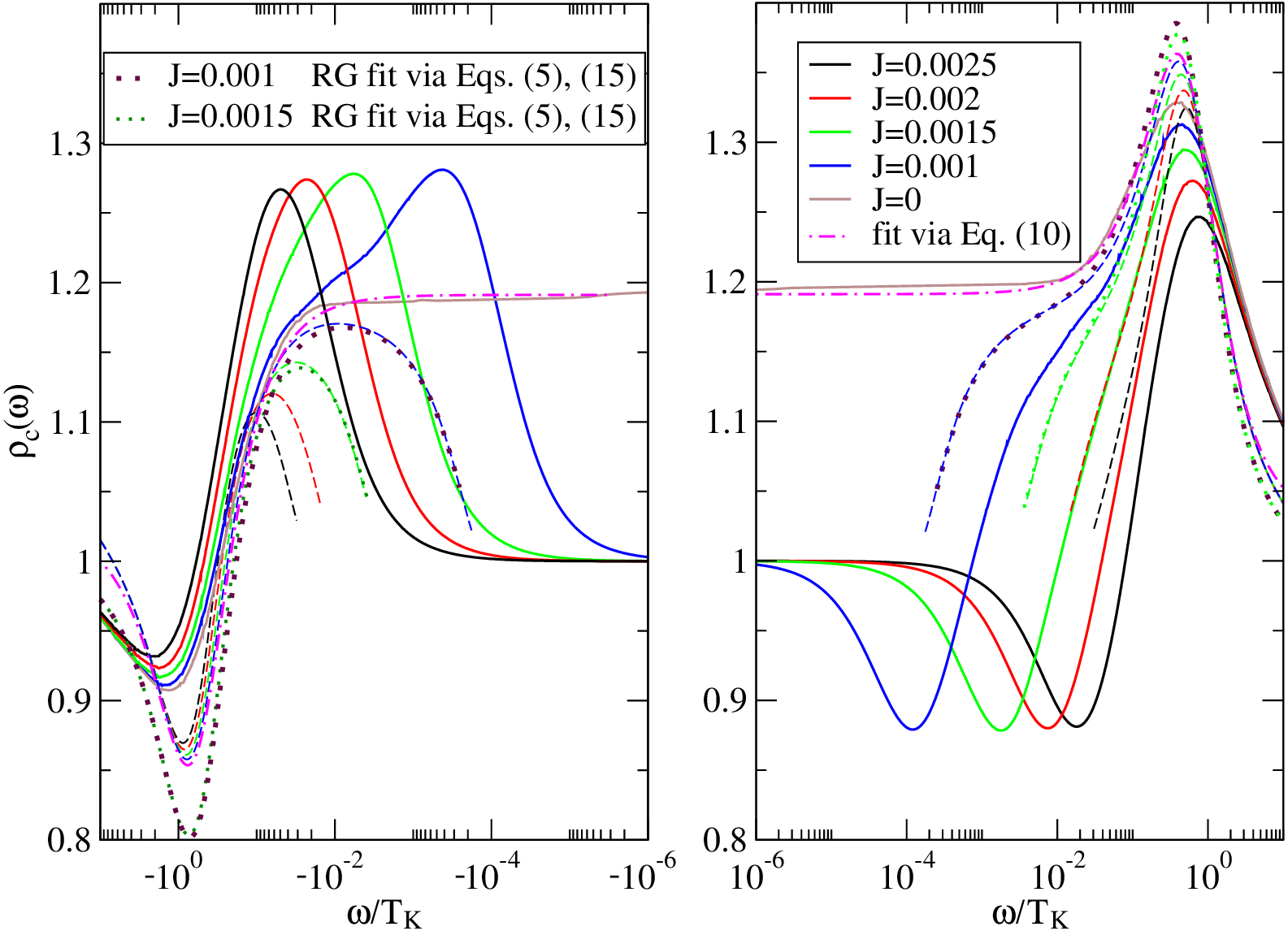}
\end{center}
\par
\vspace{-0.5cm}
\caption{The Fano lineshape for $\rho_c(\omega)$ 
(in unit of $\rho_0$) on a logarithmic scale of $\omega/T_k$ 
for different antiferromagnetic RKKY couplings $J$ 
by NRG (solid lines) and perturbative RG via Eqs. \ref{Gd1J0} and \ref{G_d1J} 
(dashed lines).  The dot-dashed lines are fits to the Fano lineshape 
form via Eq. \ref{Fano_form} for $J=0$. The dotted lines 
are the RG fits via Eqs. \ref{Gd10} and \ref{G_d1J}. 
The other parameters are the same as in Fig. 1.}
\vspace{0.3cm}
\label{Fano_log}
\end{figure}

\begin{figure}[t]
\begin{center}
\includegraphics[width=8.5cm]{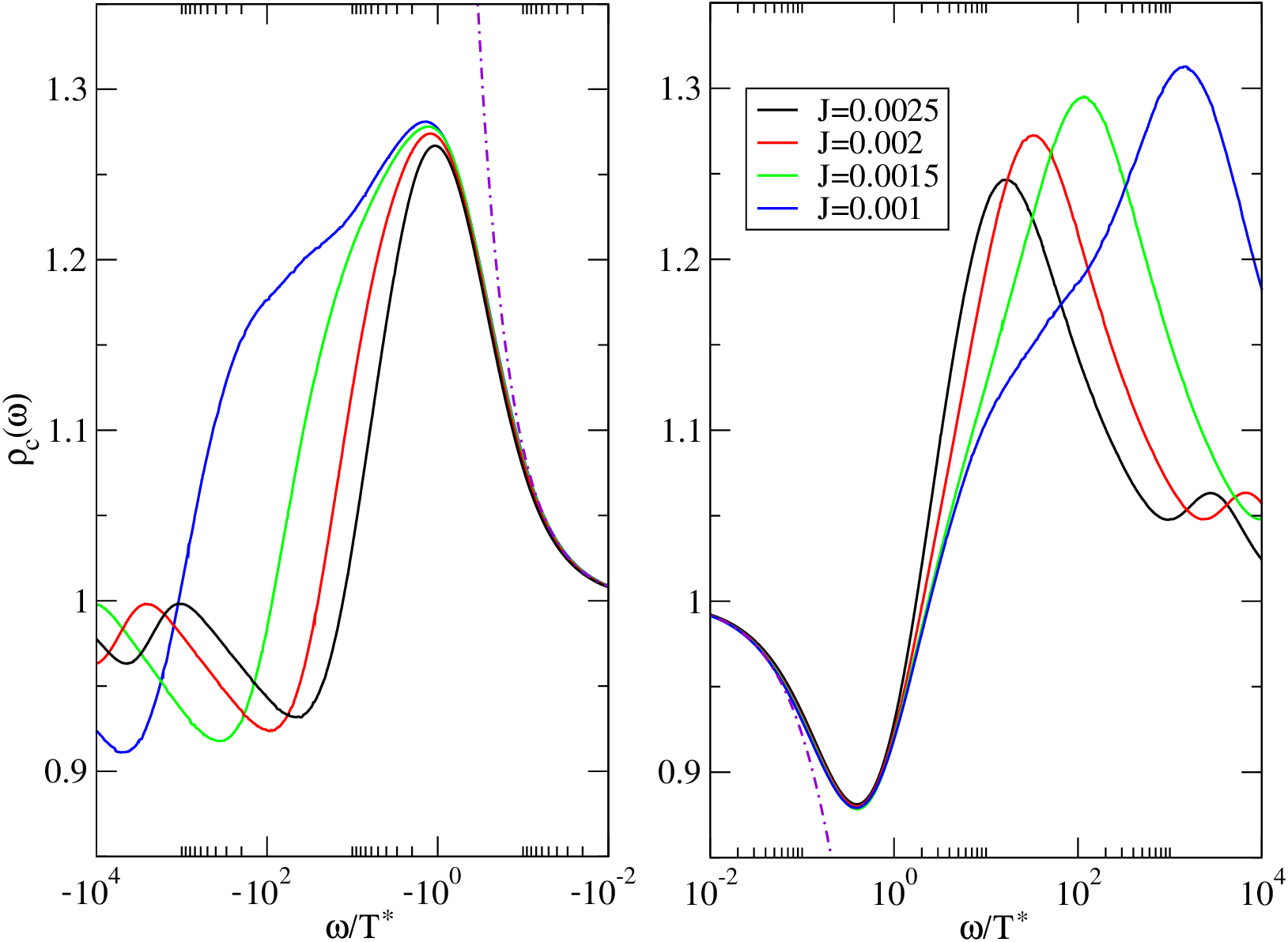}
\end{center}
\par
\vspace{-0.5cm}
\caption{The Fano lineshape for $\rho_c(\omega)$ 
(in unit of $\rho_0$) on a logarithmic scale of $\omega/T^*$ 
for different antiferromagnetic RKKY couplings $J$ 
by NRG. The other parameters are the same as in Fig. 1. 
The dot-dash lines are power-law fits to 
the universal scaling function of $\omega/T^*$ via Eq. \ref{Fano-power-law}.}
\vspace{0.3cm}
\label{Fano_scaling}
\end{figure}

From the perturbative RG results, 
as $\omega\rightarrow T^*$, $\tilde{J}(\omega)$ 
diverges, leading to the vanishing LDOS. 
As $\omega$ decreases to $T^*$ from above, the factor 
$1-  \frac{3 \tilde{J}^2(\omega) }{16 \tilde{T}_k^2}$ 
in Eq. \ref{analyticReGd} 
first becomes negative then it approaches $0$ as $\omega$ further 
approaches $0$. This explains the additional dip-peak structure seen for 
$|\omega| \rightarrow T^*$ in the NRG results. 
This qualitative feature can be captured by the perturbative RG approach. 
However, the magnitudes of the dip-peak features via  
perturbative RG are much smaller than those obtained from NRG. 
We believe the reasons for the deviation 
are two folds: First, the overall shape of $ReG_{d1}(\omega)$ predicted via 
RG is shifted towards the smaller $|\omega|$ region compared to the NRG  
results. This leads to a smaller value for $\omega_0 >0$ (compared to that 
via NRG) where $ReG_{d1}(|\omega| < \omega_0)$ changes its sign 
from positive (negative) to negative (positive) for $0< \omega<\omega_0$ 
($-\omega_0<\omega<0$). This makes the magnitudes of these 
additional dips and peaks smaller as $\tilde{J}(\omega)$ diverges 
even further (see Eq. \ref{analyticReGd}).  
As $J$ is further increased, 
the deviations between RG and NRG become more transparent. 
This is expected as the perturbation theory becomes 
uncontrolled once the system moves away from the weak coupling 
regime. Nevertheless, the perturbative RG approach can still capture 
the qualitative features of $ReG_{d1}(\omega)$ 
for $T^*<|\omega| <T_k$ (see Fig. \ref{ReGd} and Fig. \ref{ReGd_log}). 
Note that the perturbative RG approach via Eq. \ref{Gd1J0} 
(the Dyson's equation)  
can fit the NRG result  
for $ReG_{d1}(\omega)$ better than that via  
Eq. \ref{Gd10} for $\omega \ge T_k$, as expected. 
Similar to 
the KT scaling behavior for $ImG_{d1}(\omega)$, the NRG results for 
$ReG_{d1}(\omega)$ also show a scaling behavior for $\omega\le T^*$: 
$ReG_{d1}(\omega)\approx g_0^{\prime} g^{\prime}(\omega/T^*)$ 
(see Fig. \ref{ReGd_scaling}). Here, the scaling function 
$g^{\prime}(\omega/T^*)$ is again completely universal in the 
case of particle-hole symmetry.
Based on the Fermi liquid theory, the very low energy ($\omega\ll T^*$) 
crossover function for $ReG_{d1}(\omega)$ 
is linear in $\omega/T^*$ 
(see, for example Eq. \ref{ReImGdapprox}):
\begin{equation}
\Gamma ReG_{d1}(\omega) \approx - a_2 (\frac{\omega}{T^*}).
\label{regd_scale}
\end{equation} 
where we find $a_2\approx 1.5$ from the fit to the NRG result (see Fig. 
\ref{ReGd_scaling}). \\

Finally, we discuss the behavior for the Fano lineshape for $\rho_c(\omega)$. 
As indicated in Eq. \ref{Fano_form_J}, the Fano lineshape for $\delta 
\rho_c(\omega)$ 
is effectively a linear combination of the asymmetric real part 
and symmetric imaginary part of the 
$G_{d1}(\omega)$. The parameter $q_c$ in Eq. \ref{Fano_form_J}
depends on the conduction electron reservoir. 
Following Ref. \cite{kroha} and Ref. \cite{xiang}, 
$q_c$ can be reasonably treated as a constant. We take a realistic 
value $q_c\approx 1.4$ here, corresponding to the $Co/Au$ 
system studied in Ref. \cite{xiang} and Ref. \cite{wingreen}. 
We find $\rho_c(\omega)$ is asymmetric 
with respect to $\omega=0$ with a larger magnitude for $\omega>0$ than 
that for $\omega<0$. As shown in Fig. \ref{Fano_log} and  
Fig. \ref{Fano_scaling}, 
$\rho_c(\omega)$ shows a 
dips at $\omega \approx -T_k$ and $\omega \approx T^*$ as well as peaks  
at $\omega\approx T_k$ and  $\omega\approx -T^*$. The peak (dip) at 
$\omega \approx \pm T_k$ correspond to the first stage Kondo effect; while 
the dip (peak) at $\omega \approx \pm T^*$ correspond to the 
second stage Kondo effect via RKKY coupling.  
We find a reasonably good agreement 
between the NRG results and the fit via 
the perturbative RG approach for $T_k\approx \omega<T^*$. 
(The fit via Eq. \ref{Gd1J0} is somewhat better than that via Eq. \ref{Gd10} 
as the former gives a better fit to the NRG result for $ReG_{d1}(\omega)$ 
The above dip-peak structure in the Fano lineshpe for $\rho_c(\omega)$ 
in the presence of RKKY coupling 
can be detected in the STM measurement of the conduction electron 
leads as the signature of the two-stage Kondo effect in side-coupled 
double quantum dot. Note that the $\omega/T^*$ scaling in 
the NRG results for $\rho_c(\omega)$ is observed (see Fig. \ref{Fano}, 
and Fig. \ref{Fano_log}), 
which comes naturally from the scaling behaviors for both real 
and imaginary parts of $G_{d1}(\omega)$ (see Eq. \ref{deltarhoc} and 
Fig. \ref{Fano_scaling}). In the low energy limit $\omega \ll T^*$ 
where the system approaches to the Fermi-liquid of local spin singlet, 
we have the following approximated power-law 
scaling behavior for $\delta \rho_c(\omega)$: 
\begin{equation}
\frac{\delta \rho_c(\omega)}{\rho_0} \approx 
-[(1-q_c^2) a_1 (\frac{\omega}{T^*})^{2}
+ 2 a_2 q_c \frac{\omega}{T^*}]. 
\label{Fano-power-law}
\end{equation}
We would like to make one side remark here. 
For $\omega \ll T_k$, the single Lorentzian 
approximation Eq. \ref{Gd10} can very well describe 
$G_{d1}(\omega)$; however, 
for $\omega \le T_k$, we expect 
a finite contribution to $G_{d1}(\omega)$ 
from interference between 
the Kondo resonance and the broadened impurity level at dot $1$. 
We find indeed a better agreement between 
the analytic fits  and the NRG results for 
$G_{d1}(\omega)$ and $\rho_c(\omega)$ 
via perturbative 
RG approach based on the Dyson's equation Eq. \ref{Gd1J0} than 
those from the single Lorentzian fit via Eq. \ref{Gd10}.\\
\begin{center}
{\bf V. Conclusions.} 
\end{center}
We have studied the Fano resonance in a side-coupled 
double-quantum-dot system 
in the Kondo regime in the presence of particle-hole symmetry. 
In the range where 
the energy of the dot $1$ is of the order of the broadening of its  
energy level, quantum interference between the Kondo effect and 
the broadened energy level of the dot $1$ gives rise to modification 
of the Green's function on dot $1$. We apply the perturbative and numerical 
renormalization group approaches to describe the Fano lineshape in TDOS 
of the conduction electrons, which depend on both the real and imaginary 
parts of the Green's function $G_{d1}(\omega)$ of the dot $1$. At $J=0$, 
$ImG_{d1}(\omega)$ shows the Kondo peak for $\omega\le T_k$; while 
$ReG_{d1}(\omega)$ exhibits a peak (dip) for $\omega\approx T_k$ ($\omega\approx -T_k$). As a result of the Kondo effect, 
the Fano lineshape in TDOS of the conduction electron leads   
 shows a peak (dip) around $\omega\approx T_k$ 
($\omega\approx - T_k$).   
At a finite antiferromagnetic 
spin exchange coupling between the two dots, 
the two-stage Kondo effect leads to the suppression of the density of states 
on dot $1$ as well as an additional 
dip (peak) structure in the real part of $G_{d1}(\omega)$ at 
$\omega \approx \pm T^*$ from the NRG results.   
This leads to an additional dip (peak) around $\omega\approx T^*$ 
($\omega\approx -T^*$)   
in the conduction electron LDOS. 
The spliting between dip and peak in LDOS at $\omega\approx \pm T^*$ 
becomes more pronounced as the RKKY coupling $J$ is increased.  
At finite values of $J$ and for $\omega< T^*$,  
the NRG results for 
$ReG_{d1}(\omega)$, $ImG_{d1}(\omega)$ and $\rho_c(\omega)$ all 
show distinct universal scaling behaviors in $\omega/T^*$. 
 Analytically, we find the perturbative RG approach 
can qualitatively capture the above behaviors for $T^*\ll\omega \le T_k$. 
In particular, compare to the simple Lorentzian approximation 
for $G_{d1}(\omega)$, we find a better fit to the NRG results for the 
Fano lineshape for $\rho_c(\omega)$ for 
$T^*\ll \omega \le T_k$ by taking into 
account the interference between the Kondo resonance and the broadened 
impurity level on dot $1$ within the Dyson's equation approach. 
To make contact of our results in the experiments, 
the asymmetrical double dip/peak structure and the scaling behaviors 
in the Fano lineshape predicted here 
in the spectral properties of the 
TDOS of the conduction electron leads can be dectcted 
by the transport through the STM tips\cite{kroha} 
as an indication and direct consequence of the two-stage Kondo effect in 
our side-coupled double-quantum-dot system. 
Finally, we would like to make a remark on the Fano lineshape in TDOS 
of the leads in our system without  
particle-hole symmetry. In this case, we expect a smooth crossover 
(instead of the KT type transition)  
between the Kondo and local singlet phases due to the  
potential scattering terms generated in the presence of particle-hole 
asymmetry. Nevertheless, further investigations via NRG 
are needed to clarify this issue.

\acknowledgements

We are grateful for the useful discussions with Tao Xiang and  
P. W\"olfle.  We also 
acknowledge the generous support from the NSC grant 
No.95-2112-M-009-049-MY3, 98-2918-I-009-006, 98-2112-M-009-010-MY3, 
the MOE-ATU program, the NCTS of Taiwan, R.O.C., 
and National Center for Theoretical Sciences (NCTS) of Taiwan.

\end{document}